\documentclass[journal,10pt,twocolumn,twoside]{IEEEtran}

\IEEEoverridecommandlockouts
\usepackage{amsmath}
\usepackage{amsfonts}
\usepackage{cases}
\usepackage{setspace}
\usepackage{fancybox}
\usepackage{subfigure}
\usepackage{epsfig}
\usepackage{graphicx}
\usepackage{epstopdf}
\usepackage{float}
\usepackage{multirow}
\usepackage{color}
\usepackage{amsmath}
\usepackage{multirow}
\usepackage{indentfirst}
\usepackage{dsfont}
\usepackage{amsfonts}
\usepackage{times,amsmath,color,amssymb,epsfig,cite,subfigure,algorithm,algorithmic}
\usepackage{balance}

\newtheorem{theorem}{Theorem}

\usepackage{color}

\begin{document}
\title{On the design of Massive MIMO-QAM detector via $\ell_2$-Box ADMM approach}
\author{Jiangtao Wang, Quan Zhang, Yongchao Wang}

\markboth{}
{}

\maketitle

\begin{abstract}
In this letter, we develop an $\ell_2$-box maximum likelihood (ML) formulation for massive multiple-input multiple-output (MIMO) quadrature amplitude modulation (QAM) signal detection and customize an alternating direction method of multipliers (ADMM) algorithm to solve the nonconvex optimization model.
In the $\ell_2$-box ADMM implementation, all variables are solved analytically. Moreover, several theoretical results related to convergence, iteration complexity, and computational complexity are presented. Simulation results demonstrate the effectiveness of the proposed $\ell_2$-box ADMM detector in comparison with state-of-the-arts approaches.
\end{abstract}

\begin{IEEEkeywords}
 Massive MIMO, maximum-likelihood detection, $\ell_2$-box ADMM, nonconvex optimization
\end{IEEEkeywords}

\IEEEpeerreviewmaketitle

\section{Introduction}
\IEEEPARstart{M}{assive} multiple-input multiple-output (MIMO) system is a key technology for the fifth-generation (5G) wireless communication systems. It is a scaled-up version of the conventional MIMO systems that employ a large number of antenna elements at the transmitter and/or receiver for achieving higher spectral/energy efficiency \cite{albreem2019massive}.

How to detect the received signal efficiently is a well known problem in the massive MIMO system, which has attracted a large amount of attention in the field of digital communications. The maximum likelihood (ML) detector and its variants, such as sphere decoding (SD) \cite{damen2003maximum} and K-best when K is large enough \cite{Wong2002A}, can obtain optimal detection performance, but suffer from exponentially increasing computational complexity with the number of antennas.
The semidefinite relaxation (SDR) \cite{Zhi2010Semidefinite} and triangular approximate semidefinite relaxation (TASER) \cite{castaneda2016data} can achieve near-optimal detection performance with a polynomial computational complexity. However, their computational complexity is still unaffordable in massive MIMO systems with high-order modulation signals.
For a classical linear detector, such as minimum mean square error (MMSE) \cite{6771364}, it has a lower computational complexity, which comes at the expense of the loss of detection performance.

In recent years, the alternating direction method of multipliers (ADMM) technique was applied to MIMO detection due to its simplicity, operator splitting capability, and guaranteed-convergence. Although ADMM-like detectors \cite{shahabuddin2021admm, 7526551, ZhangTWC2022} can achieve competitive detection performance compared to conventional ones, their detection performance could be improved further by tightening the constraints set of the optimization model. Following this idea, in this letter, we propose a novel detector for quadrature amplitude modulation (QAM) massive MIMO systems, which exploits the $\ell_2$-box intersection technique \cite{wu2019ell} to transform the high-order QAM detection problem to a binary optimization one, and then a customized ADMM solving algorithm with theoretically-guaranteed performance, called $\ell_2$-box ADMM, is proposed. Meanwhile, several theoretical results on the proposed  $\ell_2$-box ADMM detector, related to convergence, iteration complexity, and computational complexity are presented. Simulation results demonstrate its competitive detection performance in comparison with the state-of-the-art detectors.

\section{Preliminaries}\label{sec:problem-formulation}

Consider ML detection of the uncoded signal lying in the uplink of a massive MIMO system with $U$ transmit antennas and $B$ receive antennas ($ B \ge U$). In particular, we have the following transmission model
\begin{equation}\label{transmission model}
\mathbf {r}_c = \mathbf {H}_c\mathbf {x}_c + \mathbf {n}_c,
\end{equation}
where $\mathbf{x}_c\in \mathbb{X}_c^{U}$ is the transmitted signal vector and $\mathbb{X}_c$ refers to the signal constellations set, $\mathbf{r}_c\in \mathbb{C}^{B}$ is the received signal vector, $\mathbf{H}_c\in \mathbb{C}^{B\times U}$ denotes the MIMO channel matrix, and $\mathbf{n}_c\in \mathbb{C}^{B}$ denotes additive white Gaussian noise. The entries of $\mathbf{H}_c $ and $\mathbf{n}_c $ are assumed to be independent and identically distributed (i.i.d.) complex Gaussian variables with zero mean.

To facilitate discussion later, we re-parameterize \eqref{transmission model}  using real valued vectors as described below
\begin{equation}\label{transmission model_real}
\mathbf {r} = \mathbf {H}\mathbf {x} + \mathbf {n}.
\end{equation}
In \eqref{transmission model_real}, we use the following conventions
\begin{equation}\label{convention}
\begin{split}
&\mathbf {r}= {\left[\begin{array}{c} \Re (\mathbf {r}_c) \\ \Im (\mathbf {r}_c) \end{array}\right]},
\mathbf {x}= {\left[\begin{array}{c} \Re (\mathbf {x}_c) \\ \Im (\mathbf {x}_c) \end{array}\right]},
\mathbf {n}= {\left[\begin{array}{c} \Re (\mathbf {n}_c) \\\Im (\mathbf {n}_c) \end{array}\right]}, \\
&\mathbf {H}= {\left[\begin{array}{cc} \Re (\mathbf {H}_c) & -\Im (\mathbf {H}_c)\\ \Im (\mathbf {H}_c) & \Re (\mathbf {H}_c) \end{array}\right]},
 \end{split}
\end{equation}
where $\Re(\cdot)$ and $\Im(\cdot)$ denote the real and imaginary parts of the complex variable respectively. Then, for a $4^Q$-QAM MIMO system, the optimal ML detector can be formulated as the following discrete least square problem
\begin{equation}\label{eq:MLdetection}
\min_{\mathbf{x}\in \mathbb{X}^{2U}} \Vert\mathbf{r}-\mathbf{H}\mathbf{x} \Vert_2^{2},
\end{equation}
where ${\mathbb{X}} = \{\pm 1, \pm 3, \cdots , \pm (2^{Q} -1)\}$ and $Q$ is some positive integer.
The model \eqref{eq:MLdetection} is a typical combination optimization problem since the constraint $\mathbf{x}\in \mathbb{X}^{2U}$ is discrete. It means that obtaining its global optimal solution is prohibitive in practice since the corresponding computational complexity grows exponentially with the transmit antenna number $U$, receive antenna number $B$, and the set $\mathbb{X}$'s size. In the following section, we proposed a new detector, called $\ell_2$-box ADMM, to handle this difficult problem.

\section{$\ell_2$-box ADMM detector}

\subsection{Problem formulation}
In this subsection, we propose a two-step procedure to formulate \eqref{eq:MLdetection} to a solvable optimization model using the $\ell_2$-box intersection technique.

First, we decompose vector $\mathbf{x}\in \mathbb{X}^{2U}$ into a sum of multiple binary vectors as follows
$\mathbf{x}=\textstyle\sum\nolimits_{q=1}^{Q}2^{q-1}\mathbf{x}_q$,
where $\mathbf{x}_q \in \mathbb{X}_q^{2U} = \{-1,1\}^{2U}$, $q = 1,\cdots,Q$. Plugging it into the model \eqref{eq:MLdetection}, it can be equivalent to
\begin{subequations}\label{neq:binary_MLdetection}
\begin{align}
&\hspace{0.4cm}\min_{\mathbf{x}_q} \hspace{0.2cm} \frac{1}{2}\Vert\mathbf{r}-\mathbf{H}(\textstyle\sum_{q=1}^{Q}2^{q-1}\mathbf{x}_q)\Vert_2 ^{2},  \label{eq:binary_MLdetection_a} \\
&\hspace{0.5cm}{\rm {s.t.}} \hspace{0.3cm} \mathbf{x}_q \in \mathbb{X}_q^{2U},\ q = 1,\cdots,Q. \label{eq:binary_MLdetection_b}
\end{align}
\end{subequations}

Second, replacing the binary integer constraints \eqref{eq:binary_MLdetection_b} by an intersection set between a box $\mathbb{S}_b=[-1, 1]^{2U}$ and an $\ell_2$-norm sphere $\mathbb{S}_p=\left\{\mathbf{x_q}:\Vert \mathbf{x_q} \Vert_2^2 = 2U \right\}$ \cite{wu2019ell}, we transform \eqref{neq:binary_MLdetection} to
\begin{subequations}\label{neq:binary_MLdetection2}
\begin{align}
&\hspace{0.4cm}\min_{\mathbf{x}_q} \hspace{0.2cm} \frac{1}{2}\Vert\mathbf{r}-\mathbf{H}(\textstyle\sum_{q=1}^{Q}2^{q-1}\mathbf{x}_q)\Vert_2 ^{2},  \label{eq:binary_MLdetection2_a} \\
&\hspace{0.5cm}{\rm {s.t.}} \hspace{0.3cm} \mathbf{x}_q \in \mathbb{S}_b, \ \ \mathbf{x}_q \in \mathbb{S}_p, \ q = 1,\cdots,Q. \label{eq:binary_MLdetection2_b}
\end{align}
\end{subequations}
Since constraint $\mathbf{x}_q\in\mathbb{S}_2$ is still nonconvex, solving \eqref{neq:binary_MLdetection2} directly is still difficult. In the next subsection, we develop an ADMM algorithm, which can handle $\mathbf{x}_q\in\mathbb{S}_2$ efficiently.

\subsection{$\ell_2$-box ADMM Solving Algorithm}
\label{sec:IS-ADMM Algorithm}
Introducing auxiliary variables $\mathbf{z}_{1q}$ and $\mathbf{z}_{2q}$ to the model \eqref{neq:binary_MLdetection2}, it can be equivalent to
\begin{equation}\label{eq: L2-Box ML}
\begin{split}
&\min_{\mathbf{x}_q, \mathbf{z}_{1q}, \mathbf{z}_{2q}} \frac{1}{2}\Vert\mathbf{r}\!-\!\mathbf{H}\textstyle\sum_{q=1}^{Q}2^{q-1}\mathbf{x}_q \Vert_2 ^{2}\!,\\
&\hspace{0.5cm}\text{s.t.} \hspace{0.6cm}
\mathbf{x}_q = \mathbf{z}_{1q},\quad \mathbf{x}_q = \mathbf{z}_{2q}, \\
&\hspace{1.5cm} \mathbf{z}_{1q} \in \mathbb{S}_b, \quad \mathbf{z}_{2q} \in \mathbb{S}_2, \quad q = 1,\cdots,Q.
\end{split}
\end{equation}
We write its augmented Lagrangian function\footnote{The considered augmented Lagrangian function is slightly different from the classical one. We introduce a penalty parameter for each linear constraint.} as
\begin{equation}
\begin{split}\label{eq:lagrangian_L2BoxADMM}
&L_{\rho_{1q},\rho_{2q}}(\{\mathbf{z}_{1q},\mathbf{z}_{2q},\mathbf{x}_q, \mathbf{y}_{1q},\mathbf{y}_{2q}\}_{q=1}^{Q}) \\
=& \frac{1}{2}\Vert\mathbf{r}-\mathbf{H}\textstyle\sum\nolimits_{q=1}^{Q}\!2^{q-1}\mathbf{x}_q \Vert_2 ^{2}\\
&+\textstyle\sum\nolimits_{q=1}^{Q}\mathbf{y}_{1q}^T\big(\mathbf{x}_q\!-\!\mathbf{z}_{1q}\big) \!+ \!\sum\nolimits_{q=1}^{Q}\frac{\rho_{1q}}{2}\big\|\mathbf{x}_q\!-\!\mathbf{z}_{1q}\big\|_2^2 \\
&+\textstyle\sum\nolimits_{q=1}^{Q}\mathbf{y}_{2q}^T\big(\mathbf{x}_q\!-\!\mathbf{z}_{2q}\big) + \sum\nolimits_{q=1}^{Q}\frac{\rho_{2q}}{2}\big\|\mathbf{x}_q\!-\!\mathbf{z}_{2q}\big\|_2^2,
\end{split}
\end{equation}
where $\mathbf{y}_{1q}, \mathbf{y}_{2q} \in \mathbb{R}^{2U}$ are Lagrangian multipliers, and $\rho_{1q}>0$ and $\rho_{2q}>0$ are penalty parameters.
Based on \eqref{eq:lagrangian_L2BoxADMM}, the proposed $\ell_2$-box ADMM solving algorithm framework can be described as
\begin{subequations}\label{L2BoxADMM_update}
\begin{align}
&\mathbf{z}_{1q}^{k+1} = \mathop{\arg\min}_{\mathbf{z}_{1q}\in\mathbb{S}_b}  L_{\rho_{1q},\rho_{2q}}\!(\!\{\mathbf{z}_{1q},\mathbf{z}_{2q}^{k},\!\mathbf{x}_q^k, \!\mathbf{y}_{1q}^k,\!\mathbf{y}_{2q}^k\}_{q=1}^{Q}\!), \label{z_1q_update_L2Box}\\
&\mathbf{z}_{2q}^{k+1} = \mathop{\arg\min}_{\mathbf{z}_{2q}\in\mathbb{S}_2}  L_{\rho_{1q},\rho_{2q}}\!(\!\{\mathbf{z}_{1q}^{k+1},\mathbf{z}_{2q},\!\mathbf{x}_q^k, \!\mathbf{y}_{1q}^k,\!\mathbf{y}_{2q}^k\}_{q=1}^{Q}\!), \label{z_2q_update_L2Box}\\
&\mathbf{x}_q^{k+1}\!=\! \mathop{\arg\min}_{\mathbf{x}_q} L_{\rho_{1q},\rho_{2q}}\!(\!\{\mathbf{z}_{1q}^{k+1},\!\mathbf{z}_{2q}^{k+1},\!\mathbf{x}_q, \!\mathbf{y}_{1q}^k,\!\mathbf{y}_{2q}^k\}_{q=1}^{Q}\!),\label{x_q_update_L2Box}\\
&\mathbf{y}_{1q}^{k+1}=\mathbf{y}_{1q}^k+\rho_{1q}(\mathbf{x}_q^{k+1}-\mathbf{z}_{1q}^{k+1}),\label{y_1q_update_L2Box}\\
&\mathbf{y}_{2q}^{k+1}=\mathbf{y}_{2q}^k+\rho_{2q}(\mathbf{x}_q^{k+1}-\mathbf{z}_{2q}^{k+1}).\label{y_2q_update_L2Box}
\end{align}
\end{subequations}
where $k$ is the iteration number.

\subsection{Implementation}
The main cost to implement algorithm \eqref{L2BoxADMM_update} lies in solving subproblems \eqref{z_1q_update_L2Box}--\eqref{x_q_update_L2Box}. In sequel, we show that all these subproblems can be solved efficiently by exploiting their inherent structures.

\subsubsection{Solving \eqref{z_1q_update_L2Box} and \eqref{x_q_update_L2Box}}
In \eqref{z_1q_update_L2Box}, its objective function is strongly convex and variables in $\mathbf{z}_{1q}$ are separate. Therefore, its optimal solution can be determined by solving linear gradient equation $\nabla L_{\rho_{1q},\rho_{2q}}\!(\mathbf{z}_{1q}, \cdot)=0$ and projecting every entry in the solution vector onto set $[-1, 1]$, i.e.,
\begin{equation}\label{eq: solution_Y1q}
  \mathbf{z}_{1q}^{k+1} = \underset{[-1,1]}\Pi \big(\mathbf{x}_{q}^k+{\mathbf{y}_{1q}^k}/{\rho_{1q}}\big).
\end{equation}
Subproblem \eqref{x_q_update_L2Box} is an unconstrained strongly quadratic optimization model. Therefore, its optimal solution can be obtained easily by solving $\nabla L_{\rho_{1q},\rho_{2q}}\!(\mathbf{x}_{q}, \cdot)=0$, i.e.,
\begin{equation}\label{eq: solution_Xq}
 \begin{split}
  \mathbf{x}_q^{k\!+\!1}=&{\mathbf{P}}^{-1}\!\Bigg(\! 2^{q\!-\!1}\mathbf{H}^{T}\mathbf{r}\!+\!2^{q\!-\!1}\mathbf{H}^{T}\!\mathbf{H}\Big(\!\sum_{i=1}^{q\!-\!1}\!2^{i\!-\!1} \!\mathbf{x}_i^{k\!+\!1}\!\!+\!\!\!\!\sum_{i=q\!+\!1}^{Q}\!\!\!2^{i\!-\!1}\!\mathbf{x}_i^{k}\Big)\!\!\\
 &+\!\!{\rho_{1q}}\mathbf{z}_{1q}^{k\!+\!1}\!\!+\!\!{\rho_{2q}}\mathbf{z}_{2q}^{k\!+\!1}\!\!-\!\!\mathbf{y}_{1q}^{k}\!\!-\!\!\mathbf{y}_{2q}^{k} \!\Bigg),
 \end{split}
\end{equation}
where $\mathbf{P} = 4^{q-1}\mathbf{H}^{T} \mathbf{H}+(\rho_{1q}+\rho_{2q})\mathbf{I}$ and $\mathbf{I}$ is an identity matrix.

\subsubsection{Solving \eqref{z_2q_update_L2Box}}
We rewrite \eqref{z_2q_update_L2Box} as
\begin{subequations}\label{neq:z_2q}
\begin{align}
&\hspace{0.4cm}\min_{\mathbf{z}_{2q}} \hspace{0.2cm} \frac{\rho_{2q}}{2}\big\|\mathbf{x}^k_q\!-\!\mathbf{z}_{2q}\big\|_2^2+\mathbf{y}_{2q}^{kT}\big(\mathbf{x}^k_q\!-\!\mathbf{z}_{2q}\big),  \label{neq:z_2qa} \\
&\hspace{0.5cm}{\rm {s.t.}} \hspace{0.3cm} \|\mathbf{z}_{2q}\|_2^2=2U, \label{neq:z_2qb}
\end{align}
\end{subequations}
which can be further equivalent to
\begin{subequations}\label{neq:z_2q2}
\begin{align}
&\hspace{0.4cm}\min_{\mathbf{z}_{2q}} \hspace{0.2cm} -(\rho_{2q}\mathbf{x}^{k}_q+\mathbf{y}^k_{2q})^T\mathbf{z}_{2q},  \label{neq:z_2qa2} \\
&\hspace{0.5cm}{\rm {s.t.}} \hspace{0.3cm} \|\mathbf{z}_{2q}\|_2^2=2U. \label{neq:z_2qb2}
\end{align}
\end{subequations}
According to triangular inequality, its optimal solution $\mathbf{z}_{2q}^{k+1}$ should satisfy
\begin{equation}\label{neq:z_2q3}
\mathbf{z}_{2q}^{k+1}=c(\rho_{2q}\mathbf{x}^{k}_q+\mathbf{y}^k_{2q}),
\end{equation}
where $c$ is a positive constant to let $\mathbf{z}_{2q}^{k+1}$ satisfy \eqref{neq:z_2qb2}. Plugging \eqref{neq:z_2q3} into $ \|\mathbf{z}_{2q}\|_2^2=2U$, we can obtain
\begin{equation}\label{c}
  c = \frac{\sqrt{2U}}{\|\rho_{2q}\mathbf{x}^{k}_q+\mathbf{y}^k_{2q}\|_2},
\end{equation}
which results in
\begin{equation}\label{z_2q_k+1_2}
  \mathbf{z}_{2q}^{k+1}=\frac{\sqrt{2U}(\rho_{2q}\mathbf{x}^k_q+\mathbf{y}^k_{2q})}{\lVert\rho_{2q}\mathbf{x}^k_q+\mathbf{y}^k_{2q}\rVert_2}.
\end{equation}

{\it Remarks:} Variables $\{\mathbf{z}_{1q}|q=1,\dotsb,Q\}$ are decoupled. This means that they can be updated in parallel. Similarly, variables $\{\mathbf{z}_{2q}|q=1,\dotsb,Q\}$ are also decoupled. Thus, they can be updated in parallel too. Moreover, variables $\mathbf{z}_{1q}$ and $\mathbf{z}_{2q}$ are also separate. This means that \eqref{eq: solution_Y1q} and \eqref{z_2q_k+1_2} can be implemented in parallel. Furthermore, since variables $\{\mathbf{x}_{q}|q=1,\dotsb,Q\}$ are coupled, we have to update them in series. To be clear, we summarize the proposed $\ell_2$-box ADMM algorithm in \emph{Algorithm \ref{L2Box-ADMM algorithm}}.

\begin{algorithm}[!t]
\caption{The proposed $\ell_2$-box ADMM algorithm}
\label{L2Box-ADMM algorithm}
\begin{algorithmic}[1]
\REQUIRE $\mathbf{H}$, $\mathbf{r}$, ${Q}$, $\{\rho_{1q},\rho_{2q}\}_{q=1}^{Q}$\\
\ENSURE $\textstyle\sum\nolimits_{q=1}^{Q}2^{q-1}\mathbf{x}_q^k$\\
\STATE Initialize $\{\mathbf{z}_{1q}^{1},\mathbf{z}_{2q}^{1},\mathbf{x}_q^{1},\mathbf{y}_{1q}^1,\mathbf{y}_{2q}^1\}_{q=1}^{Q}$ as the all-zeros vectors.
\STATE  \textbf{For $k = 1,2,\cdots$}
\STATE \hspace{0.2cm} Step 1: Update $\{\mathbf{z}_{1q}^{k+1},\mathbf{z}_{2q}^{k+1}\}_{q=1}^{Q}$ in parallel via \eqref{eq: solution_Y1q} \\
\hspace{1.4cm} and \eqref{z_2q_k+1_2}, respectively.
\STATE \hspace{0.2cm} Step 2: Update $\{\mathbf{x}_q^{k+1}\}_{q=1}^{Q}$ sequentially via \eqref{eq: solution_Xq}.
\STATE \hspace{0.2cm} Step 3: Update $\{\mathbf{y}_{1q}^{k+1},\mathbf{y}_{2q}^{k+1}\}_{q=1}^{Q}$ in parallel via \eqref{y_1q_update_L2Box} \\
\hspace{1.4cm} and \eqref{y_2q_update_L2Box}, respectively.
\STATE \textbf{Until} some preset condition is satisfied.
\end{algorithmic}
\end{algorithm}

\section{Performance Analysis}\label{sec:Analysis-L2Box-ADMM}
In this section, several theoretical results of the proposed $\ell_2$-box ADMM algorithm on convergence, iteration complexity, and computational cost are presented. Here, $\bar{\mathbf{x}}_q=[\mathbf{x}_q;\mathbf{x}_q]$, $\mathbf{z}_q=[\mathbf{z}_{1q};\mathbf{z}_{2q}]$, and $\mathbf{y}_q=[\mathbf{y}_{1q};\mathbf{y}_{2q}]$, $\rho_q\!=\![\rho_{1q};\rho_{2q}]$; $\mathit{f}\left(\bar{\mathbf{x}}^*_q\right)=\frac{1}{2}\Vert\mathbf{r}-\mathbf{H}\sum\nolimits_{q=1}^{Q}\!\!2^{q-1}\bar{\mathbf{x}}^*_q \Vert_2 ^{2}$; and $\lambda_{\rm min}(\cdot)$ and $\lambda_{\rm max}(\cdot)$ denote the minimum and maximum eigenvalues of a matrix respectively. Moreover, we note that the proofs of these theoretical results are similar to a previous work \cite{ZhangTWC2022}. Due to the limited space, the details of the proof are included in the \emph{supplementary material} \cite{ZhangarXiv20220822}.

\subsection{Convergence}
We have the following theorem to characterize the convergence property of the proposed $\ell_2$-box ADMM detection algorithm.
\begin{theorem}\label{thm:convergence}
Assume $\rho_q>4^{q\!-\!1}\sqrt{2}\lambda_{\rm max}(\mathbf{H}^H\mathbf{H})$,  $q = 1,\cdots,Q$. Then, tuples $\{\{\mathbf{z}_q, \bar{\mathbf{x}}_q,\mathbf{y}_q\}_{q=1}^{Q}\}$ generated by \emph{Algorithm \ref{L2Box-ADMM algorithm}} are convergent, i.e.,
\begin{equation}\label{convergence variables}
\begin{split}
&\lim\limits_{k\rightarrow+\infty}\mathbf{z}^{k}_q=\mathbf{z}^*_q, \ \ \lim\limits_{k\rightarrow+\infty}\bar{\mathbf{x}}_q^{k}=\bar{\mathbf{x}}^*, \lim\limits_{k\rightarrow+\infty}\mathbf{y}_q^{k}=\mathbf{y}_q^*.
\end{split}
\end{equation}
Moreover, $(\{\mathbf{z}_q^*, \bar{\mathbf{x}}_q^*,\mathbf{y}_q^*\}_{q=1}^{Q}\})$ is some stationary point of the problem \eqref{eq: L2-Box ML} in the sense that
\begin{equation}\label{stationary point}
\begin{split}
&\big \langle\mathbf{z}_{1q}-\mathbf{z}^*_{1q},\  -\mathbf{y}_{1q}^*\big\rangle \ge 0,\\
&\mathbf{z}_{2q}^* \in \mathop{\arg\min}_{\mathbf{z}_{2q}\in\mathbb{S}_2}  \big \langle-\mathbf{y}_{2q}^*- {\rho_{2q}}\big(\mathbf{x}_q^*-\mathbf{z}_{2q})\big\rangle,\\
&\mathbf{y}_q^*=-\nabla_{\mathbf{x}_q} \mathit{f}\left(\bar{\mathbf{x}}_q^*\right), \quad  \bar{\mathbf{x}}_q^* = \mathbf{z}_{q}^*.
\end{split}
\end{equation}
\end{theorem}

{\it Remarks:} Theorem \ref{thm:convergence} indicates that the proposed $\ell_2$-box ADMM algorithm is theoretically-guaranteed convergent to some stationary point of the model \eqref{eq: L2-Box ML} under some mild conditions.

\subsection{Iteration complexity}
The following theorem shows the iteration complexity of the proposed $\ell_2$-box ADMM detection algorithm.
\begin{theorem}\label{iteration complexity}
  Let $t$ be the minimum iteration index such that the residual $\sum_{q=1}^{Q}\|\bar{\mathbf{x}}_q^{k+1}-\bar{\mathbf{x}}_q^{k}\|_2^2<\epsilon$, where $\epsilon$ is a desired precise parameter for the solution. Then, we have the following iteration complexity result
 \[
   \begin{split}
     t \!\leq \!\frac{1}{C\epsilon}\Big(\!L_{\rho}(\{\mathbf{z}^{1}_q, \bar{\mathbf{x}}_q^{1}, \mathbf{y}_q^{1}\}_{q=1}^{Q}) -\mathit{f}\left(\bar{\mathbf{x}}_q^*\right) \Big),
   \end{split}
 \]
where $ C\!=\!\min\left\{ \frac{\!\rho_q + 4^{q-1}\lambda_{\rm min}(\mathbf{H}^{H} \mathbf{H})}{2}\!-\!\frac{16^{q-1}\lambda_{\rm max}^2(\mathbf{H}^{H} \mathbf{H})}{\rho_q}\!\right\}_{q=1}^{Q}$.
\end{theorem}

{\it Remarks:}
Theorem \ref{iteration complexity} indicates that the ``{\it minimum iterations}'' is upper-bounded for the given precise $\epsilon$, which gives a low bound for the computational complexity of the proposed $\ell_2$-box ADMM algorithm.

\subsection{Computational cost}
Now, we analyze the computational cost of the proposed $\ell_2$-box ADMM detection algorithm. Before starting iterations, we should compute ${\Big(\!4^{q\!-\!1}\mathbf{H}^{H}\! \mathbf{H}\!\!+\!\!(\rho_{1q}\!\!+\!\!\rho_{2q})\mathbf{I}\!\Big)}\!^{-1}$ and $\mathbf{H}^{H}\!\mathbf{r}$ in \eqref{eq: solution_Xq}, which roughly needs $\mathcal{O}(U^3+BU^2+BU)$ multiplications. When implementing the algorithm, we need to perform roughly ${\mathcal{O}(Q(U^2+U))}$ multiplications in each iteration. Hence, the total computational cost is $\mathcal{O}(U^3+BU^2+BU+QT(U^2+U))$, where $T$ is the pre-set maximum number of iterations. In Table \ref{Complexity-Comparison}, we compare the computational complexity of the $\ell_2$-box ADMM algorithm with several state-of-the-art massive MIMO detection ones. We can see that the proposed $\ell_2$-box ADMM algorithm is cheaper than TASER and K-best, comparable to ADMM-based algorithms, but higher than the linear MMSE algorithm. Moreover, we should note, though OCD-BOX has the lowest complexity of all detectors, its detection efficiency is low since the variables have to be updated in series.

\begin{table}\small
\caption{Computational Complexity for Different Algorithms}
\label{Complexity-Comparison}
\begin{center}
\renewcommand\arraystretch{1.2}
\setlength{\tabcolsep}{1.3mm}{
\begin{tabular}{|c|l|}
\hline
\hline
\textbf{Algorithm}&  \textbf{Computational complexity}\\\hline
MMSE \cite{6771364} & $\mathcal{O}(U^3+BU^2+BU)$ \\\hline
OCD-BOX \cite{wu2016high}  & $\mathcal{O}(T(2BU+U))$ \\\hline
TASER \cite{castaneda2016data} & $\mathcal{O}(T(\frac{2}{3}U^3+\frac{5}{2}U^2+\frac{37}{12}U))$ \\\hline
K-best \cite{Wong2002A} & $\mathcal{O}(U^3)\sim\mathcal{O}(4^{QU})$ \\\hline
ADMM \cite{7526551} & $\mathcal{O}(U^3+BU^2+BU+T(U^2+\frac{1}{2}U))$ \\\hline
ADMIN \cite{shahabuddin2021admm} & $\mathcal{O}(U^3+BU^2+BU+T(U^2+U))$ \\\hline
PS-ADMM \cite{ZhangTWC2022} & $\mathcal{O}(U^3\!+\!BU^2\!+\!BU\!+\!T(U^2\!+\!\frac{1}{2}Q(Q\!+\!1)U))$ \\\hline
$\ell_2$-box ADMM &  $\mathcal{O}(U^3+BU^2+BU+QT(U^2+U))$ \\\hline\hline
\end{tabular}}
\end{center}
\end{table}

\section{Simulation Results}\label{sec:Simulation Results}
In this section, numerical examples are presented to show detection performance, computational complexity, and convergence characteristic of the proposed $\ell_2$-box ADMM detector compared with several state-of-the-art massive MIMO detectors, where the 16-QAM modulation scheme is considered.

\begin{figure}[htpb]
  \begin{minipage}{7.8cm}
    \centering
         \includegraphics[width=3.5in,height=2.7in]{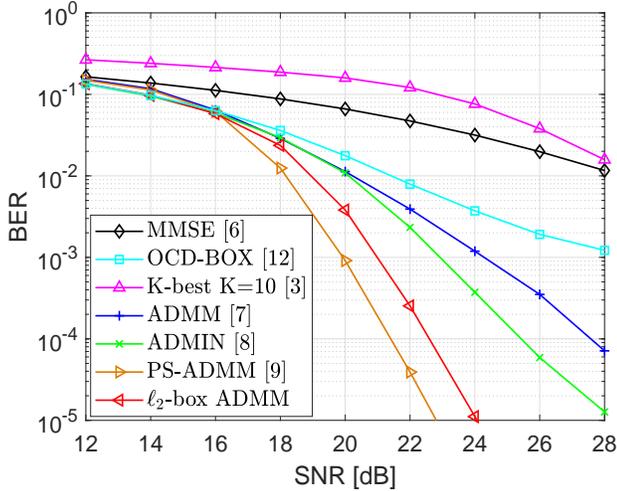}
    \caption{BER performance comparison of massive MIMO detectors with $B=128,U=128$.}
    \label{ber_allalgo}
  \end{minipage}
 \end{figure}

Fig.\,\ref{ber_allalgo} compares bit error rate (BER) performance of the different detectors for a large-scale $128\times128$ massive MIMO system. One can find that the proposed $\ell_2$-box ADMM detector achieves competitive detection performance, which outperforms most detectors, and is comparable to the PS-ADMM detector. Here, the BER curve of the TASER detector is not shown since it does not support higher-order modulation schemes.

 \begin{figure}[htpb]
  \begin{minipage}{7.8cm}
    \centering
         \includegraphics[width=3.5in,height=2.7in]{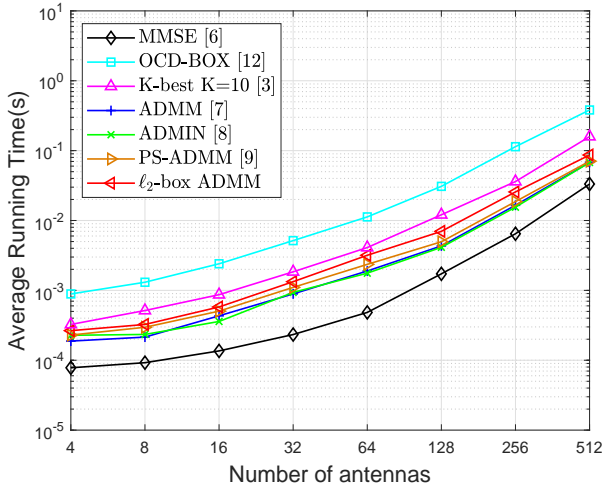}
    \caption{Complexity comparison of massive MIMO detectors with $U=B$, SNR=20dB.}
    \label{AllAlg_TimeCompare}
  \end{minipage}
 \end{figure}

Fig. \ref{AllAlg_TimeCompare} compares the computational complexity of different massive MIMO detectors. From the figure, we can observe that the running time of the proposed $\ell_2$-box ADMM detector is lower than OCD-BOX, comparable to ADMM-based detectors and K-best (K=10), and higher than MMSE.
However, we should note that the proposed $\ell_2$-box ADMM has much better detection performance than other detectors except PS-ADMM in Fig.\,\ref{ber_allalgo}. Meanwhile, the $\ell_2$-box ADMM does not have the complicated work of adjusting the penalty parameters like PS-ADMM. Therefore, the results reveal that the proposed $\ell_2$-box ADMM detector has an attractive tradeoff between BER performance and computational complexity.
\begin{figure}[htpb]
  \begin{minipage}{7.8cm}
    \centering
         \includegraphics[width=3.5in,height=2.7in]{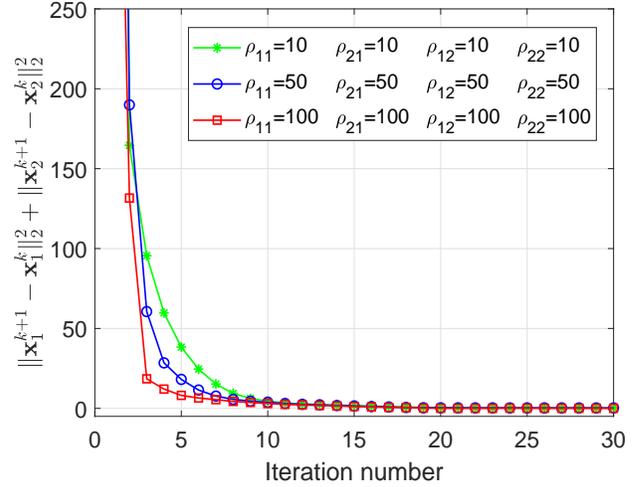}
    \caption{The impact of parameters on convergence performance of the $\ell_2$-box ADMM detector with $B=128,U=128$, SNR=20dB.}
    \label{impact_Parameters}
  \end{minipage}
\end{figure}

Fig.\,\ref{impact_Parameters} shows the impact of the parameters on the convergence characteristic of the proposed $\ell_2$-box ADMM detector.
By checking the residual $\sum_{q=1}^{2}\|\bar{\mathbf{x}}_q^{k+1}-\bar{\mathbf{x}}_q^{k}\|_2^2$,  one can find that the proposed $\ell_2$-box ADMM algorithm can always converge within 20 iterations when parameters $\rho_{1q}$ and $\rho_{2q}$, $q=1,2$ are set at different values.

\section{Conclusion}\label{sec:Conclusion}
In this letter, we proposed an efficient $\ell_2$-box ADMM detector for massive MIMO-QAM system, where all variables in the implementation can be obtained analytically. Meanwhile, we show that the ADMM iterations can converge under mild conditions. Simulation results demonstrate its effectiveness in comparison with state-of-the-arts massive MIMO detectors.

\ifCLASSOPTIONcaptionsoff
  \newpage
\fi


\end{document}


\title{On the design of Massive MIMO-QAM detector via $\ell_2$-Box ADMM approach \\ {\it Supplementary Material}}
\author{Jiangtao Wang, Quan Zhang, Yongchao Wang}

\markboth{}
{}

\maketitle

\IEEEpeerreviewmaketitle

In this section, we firstly briefly present the proposed MIMO detection optimization problem and the corresponding $\ell_2$-box ADMM algorithm in Section 1, followed by the convergence analysis and iteration complexity analysis about the $\ell_2$-box ADMM algorithm for the MIMO detection problem in Section 2 and Section 3, respectively. The part of analysis skills are motivated by \cite{hong2016convergence}.
\section{$\ell_2$-box ADMM Algorithm for the MIMO detection Problem}
\label{sec:problem-agorithm}
We formulate equivalently the maximum-likelihood MIMO detection problem to the following optimization problem
\begin{equation}\label{eq: L2-Box ML}
\begin{split}
&\min_{\mathbf{x}_q, \mathbf{z}_{1q}, \mathbf{z}_{2q}} \frac{1}{2}\Vert\mathbf{r}\!-\!\mathbf{H}\textstyle\sum_{q=1}^{Q}2^{q-1}\mathbf{x}_q \Vert_2 ^{2},\\
&\hspace{0.1cm}\text{s.t.}\begin{cases}
\mathbf{x}_q = \mathbf{z}_{1q},\quad \mathbf{x}_q = \mathbf{z}_{2q}, \quad \mathbf{x}_q \in \mathbb{R}^{2U},\\
\mathbf{z}_{1q} \in \mathbb{S}_b, \quad \mathbf{z}_{2q} \in \mathbb{S}_2, \quad q = 1,\cdots,Q,
\end{cases}
\end{split}
\end{equation}
where $\mathbb{S}_b = [-1, 1]^{2U}, \mathbb{S}_2 = \{\mathbf{z}_{2q}:\Vert \mathbf{z}_{2q} \Vert_2^{2} = 2U \}$.

The augmented Lagrangian of the problem \eqref{eq: L2-Box ML} can be expressed as
\begin{align}
\begin{split}\label{eq:lagrangian_L2BoxADMM}
&L_{\rho_{1q},\rho_{2q}}(\!\{\mathbf{z}_{1q},\!\mathbf{z}_{2q},\!\mathbf{x}_q, \!\mathbf{y}_{1q},\!\mathbf{y}_{2q}\!\}_{q=1}^{Q}\!)\! = \!\frac{1}{2}\Vert\mathbf{r}\!-\!\mathbf{H}\textstyle\sum\nolimits_{q=1}^{Q}\!\!2^{q-1}\mathbf{x}_q \Vert_2 ^{2}\\
&+\!\textstyle\sum\nolimits_{q=1}^{Q}\big\langle \mathbf{x}_q\!-\!\mathbf{z}_{1q}, \mathbf{y}_{1q}\big\rangle \!+ \!\sum\nolimits_{q=1}^{Q}\frac{\rho_{1q}}{2}\big\|\mathbf{x}_q\!-\!\mathbf{z}_{1q}\big\|_2^2 \\
&+\!\textstyle\sum\nolimits_{q=1}^{Q}\big\langle \mathbf{x}_q\!-\!\mathbf{z}_{2q}, \mathbf{y}_{2q}\big\rangle\! + \!\sum\nolimits_{q=1}^{Q}\frac{\rho_{2q}}{2}\big\|\mathbf{x}_q\!-\!\mathbf{z}_{2q}\big\|_2^2,
\end{split}
\end{align}
where $\mathbf{y}_{1q}, \mathbf{y}_{2q} \in \mathbb{R}^{2U}$ are the dual variables, and $\rho_{1q}$ and $\rho_{2q}$ are positive penalty parameters.

Let $\bar{\mathbf{x}}_q=[\mathbf{x}_q;\mathbf{x}_q]$, $\mathbf{z}_q=[\mathbf{z}_{1q};\mathbf{z}_{2q}]$, $\mathbf{y}_q=[\mathbf{y}_{1q};\mathbf{y}_{2q}]$, and $\rho_q=[\rho_{1q};\rho_{2q}]$, the augmented Lagrangian \eqref{eq:lagrangian_L2BoxADMM} be transformed as follows

\begin{align}
\begin{split}\label{eq:lagrangian_L2BoxADMM_transform}
&L_{\rho_q}(\{\mathbf{z}_q,\bar{\mathbf{x}}_q, \mathbf{y}_q\}_{q=1}^{Q})\! = \!\frac{1}{2}\Vert\mathbf{r}\!-\!\mathbf{H}\textstyle\sum\nolimits_{q=1}^{Q}\!\!2^{q-1}\bar{\mathbf{x}}_q \Vert_2 ^{2}+\!\textstyle\sum\nolimits_{q=1}^{Q}\big\langle \bar{\mathbf{x}}_q\!-\!\mathbf{z}_q, \mathbf{y}_q\big\rangle \!+ \!\sum\nolimits_{q=1}^{Q}\frac{\rho_q}{2}\big\|\bar{\mathbf{x}}_q\!-\!\mathbf{z}_q\big\|_2^2.
\end{split}
\end{align}

Based on the augmented Lagrangian \eqref{eq:lagrangian_L2BoxADMM_transform},  the proposed $\ell_2$-box ADMM solving algorithm framework can be described as
\begin{subequations}\label{L2BoxADMM_update}
\begin{align}
&\mathbf{z}_q^{k+1} = \mathop{\arg\min}_{\mathbf{z}_{1q}\in\mathbb{S}_b,\mathbf{z}_{2q}\in\mathbb{S}_2}  L_{\rho_q}(\{\mathbf{z}_q,\bar{\mathbf{x}}_q^k, \mathbf{y}_q^k\}_{q=1}^{Q}), \label{z_q_update_L2Box}\\
&\bar{\mathbf{x}}_q^{k+1} = \mathop{\arg\min}_{\bar{\mathbf{x}}_q} L_{\rho_q}(\{\mathbf{z}_q^{k+1},\bar{\mathbf{x}}_q, \mathbf{y}_q^k\}_{q=1}^{Q}),\label{xbar_q_update_L2Box}\\
&\mathbf{y}_q^{k+1}=\mathbf{y}_q^k+\rho_q(\bar{\mathbf{x}}_q^{k+1}-\mathbf{z}_q^{k+1}),\label{y_q_update_L2Box}
\end{align}
\end{subequations}
where $q=1,\cdots,Q,$ $k$ denotes iteration number.

\section{Convergence Analysis}
\label{sec:analysis-convergence}

We have the following theorem to show convergence properties of the proposed $\ell_2$-box ADMM algorithm.
\begin{theorem}\label{thm:convergence}
Assume $\rho_q>4^{q\!-\!1}\sqrt{2}\lambda_{\rm max}(\mathbf{H}^H\mathbf{H})$,  $q = 1,\cdots,Q$. Then, tuples $\{\{\mathbf{z}_q, \bar{\mathbf{x}}_q,\mathbf{y}_q\}_{q=1}^{Q}\}$ generated by $\ell_2$-box ADMM Algorithm are convergent, i.e.,
\begin{equation}\label{convergence variables}
\begin{split}
&\lim\limits_{k\rightarrow+\infty}\mathbf{z}^{k}_q=\mathbf{z}^*_q, \ \ \lim\limits_{k\rightarrow+\infty}\bar{\mathbf{x}}_q^{k}=\bar{\mathbf{x}}^*, \lim\limits_{k\rightarrow+\infty}\mathbf{y}_q^{k}=\mathbf{y}_q^*, \\
& \hspace{0.3cm} \forall~\mathbf{z}_{1q}\in\mathbb{S}_b,\mathbf{z}_{2q}\in\mathbb{S}_2, \; q = 1,\cdots,Q.
\end{split}
\end{equation}
Moreover, $(\{\mathbf{z}_q^*, \bar{\mathbf{x}}_q^*,\mathbf{y}_q^*\}_{q=1}^{Q}\})$ is some stationary point of the problem \eqref{eq: L2-Box ML} in the sense that
\begin{equation}\label{stationary point}
\begin{split}
&\big \langle\mathbf{z}_{1q}-\mathbf{z}^*_{1q},\  -\mathbf{y}_{1q}^*\big\rangle \ge 0,\\
&\mathbf{z}_{2q}^* \in \mathop{\arg\min}_{\mathbf{z}_{2q}\in\mathbb{S}_2}  \big \langle-\mathbf{y}_{2q}^*- {\rho_{2q}}\big(\mathbf{x}_q^*-\mathbf{z}_{2q})\big\rangle,\\
&\mathbf{y}_q^*=-\nabla_{\mathbf{x}_q} \mathit{f}\left(\bar{\mathbf{x}}_q^*\right), \quad  \bar{\mathbf{x}}_q^* = \mathbf{z}_{q}^*.
\end{split}
\end{equation}
\end{theorem}

{\it Remarks:}
Theorem \ref{thm:convergence} indicates that the proposed $\ell_2$-box ADMM algorithm is theoretically-guaranteed converged to some stationary point of the problem \eqref{eq: L2-Box ML} under some mild conditions. Here, we should note that these conditions are easily satisfied since the values of penalty parameters $\{\rho_q\}_{q=1}^{Q}$ can be set accordingly when the channel matrix $\mathbf{H}$ is known. The key idea of proving Theorem \ref{thm:convergence} is to find out that the potential function $L_{\rho_q}(\{\mathbf{z}_q,\bar{\mathbf{x}}_q, \mathbf{y}_q\}_{q=1}^{Q})$ {\it decreases sufficiently} in every $\ell_2$-box ADMM iteration and is lower-bounded. To reach this goal,
we first prove several related lemmas and then provide detailed proof of Theorem \ref{thm:convergence}.
For simplicity of notation, we use $\mathit{f}\left(\bar{\mathbf{x}}_q\right)$ to denote $\frac{1}{2}\Vert\mathbf{r}-\mathbf{H}\sum\nolimits_{q=1}^{Q}\!\!2^{q-1}\bar{\mathbf{x}}_q \Vert_2 ^{2}$ in the rest of this paper.

\begin{proof}
Before proving convergence of the proposed $\ell_2$-box ADMM algorithm, we give several lemmas and their proofs as follows.

\begin{lemma}\label{lemma:z1}
For Algorithm 1, the following inequality holds
\begin{equation}\label{eq:z_difference}
\|\mathbf{y}_q^{k+1}-\mathbf{y}_q^k\|_2^2 \le 16^{q-1}\lambda_{\rm max}^2(\mathbf{H}^H\mathbf{H})\|\bar{\mathbf{x}}_q^{k+1}-\bar{\mathbf{x}}_q^k\|_2^2,
\end{equation}
\end{lemma}
where $\lambda_{\rm max}(\cdot)$ denotes the maximum eigenvalues of a matrix.
\begin{proof}
Since $\bar{\mathbf{x}}_q^{k+1}$ is a minimizer of problem \eqref{xbar_q_update_L2Box}, it should satisfy the following optimality condition \cite{bertsekas2009convex}
\begin{align}\label{eq:x_nabla_expression}
\nabla_{\bar{\mathbf{x}}_q} \mathit{f}\left(\bar{\mathbf{x}}_q^{k+1}\right)+\mathbf{y}_q^k+\rho_q (\bar{\mathbf{x}}_q^{k+1}-\mathbf{z}_q^{k+1})=0.
\end{align}
Plugging $\mathbf{y}_q^{k+1}$ in \eqref{y_q_update_L2Box} into the above equation, we obtain
\begin{equation}\label{eq:z_nabla_expression}
\mathbf{y}_q^{k+1}=-\nabla_{\bar{\mathbf{x}}_q} \mathit{f}\left(\bar{\mathbf{x}}_q^{k+1}\right).
\end{equation}
According to Lagrange's mean value theorem, since $\mathit{f}\left(\bar{\mathbf{x}}_q\right)$ is continuous and differentiable, there exists some point $\tilde{\mathbf{x}}_q$ between $\bar{\mathbf{x}}_q^{k}$ and $\bar{\mathbf{x}}_q^{k+1}$ which satisfies
\begin{equation}\label{mean value theorem}
\frac{\nabla_{\bar{\mathbf{x}}_q} \mathit{f}\left(\bar{\mathbf{x}}_q^{k+1}\right)-\nabla_{\bar{\mathbf{x}}_q} \mathit{f}\left(\bar{\mathbf{x}}_q^{k}\right)}{\bar{\mathbf{x}}_q^{k+1}-\bar{\mathbf{x}}_q^{k}}= \nabla_{\tilde{\mathbf{x}}_q}^2 \mathit{f}\left(\bar{\mathbf{x}}_q\right).
\end{equation}
Moreover, since $\nabla_{\tilde{\mathbf{y}}_q}^2 \mathit{f}\left(\bar{\mathbf{x}}_q\right) =4^{q-1}\mathbf{H}^{H} \mathbf{H} \preceq 4^{q-1}\lambda_{\rm max}(\mathbf{H}^{H} \mathbf{H})\mathbf{I}$, we have \begin{equation}\label{lipschitz x0}
\begin{split}
\!\!\|\!\nabla_{\bar{\mathbf{x}}_q} \mathit{f}\left(\bar{\mathbf{x}}_q^{k+1}\right)\!\!-\!\!\nabla_{\bar{\mathbf{x}}_q} \mathit{f}\left(\bar{\mathbf{x}}_q^{k}\right)\!\!\|_2^2
 \!\!\le\!\!16^{q-1}\lambda_{\rm max}^2(\mathbf{H}^{H}\!\mathbf{H}) \|\bar{\mathbf{x}}_q^{k+1}\!\!\!-\!\bar{\mathbf{x}}_q^{k}\|_2^2.
 \end{split}
\end{equation}
From \eqref{lipschitz x0}, we can see that $\nabla_{\bar{\mathbf{x}}_q} \mathit{f}\left(\bar{\mathbf{x}}_q\right)$ is Lipschitz continuous with constant $4^{q-1}\lambda_{\rm max}(\mathbf{H}^{H} \mathbf{H})$.
Plugging \eqref{eq:z_nabla_expression} into LHS of equation \eqref{lipschitz x0}, we can obtain
\begin{align*}
\begin{split}
\|\mathbf{y}_q^{k+1}-\mathbf{y}_q^k\|_2^2
=&\|\nabla_{\bar{\mathbf{x}}_q} \mathit{f}\left(\bar{\mathbf{x}}_q^{k+1}\right)-\nabla_{\bar{\mathbf{x}}_q} \mathit{f}\left(\bar{\mathbf{x}}_q^{k}\right)\|_2^2 \le 16^{q-1} \lambda_{\rm max}^2(\mathbf{H}^{H} \mathbf{H}) \|\bar{\mathbf{x}}_q^{k+1}-\bar{\mathbf{x}}_q^{k}\|_2^2.
\end{split}
\end{align*}
This completes the proof.
\end{proof}

\begin{lemma}\label{lemma:L_difference}
For Algorithm 1, we have the following inequality
\begin{equation}\label{eq:L_difference}
\begin{split}
\!\!&L_{\rho_q}\left(\{\mathbf{z}_q^{k+1}\!, \bar{\mathbf{x}}_q^{k+1}\!, \mathbf{y}_q^{k+1}\}_{q=1}^{Q}\right)\!-\!L_{\rho_q}\left (\{\mathbf{z}_q^{k}, \bar{\mathbf{x}}_q^{k}\!, \mathbf{y}_q^{k}\}_{q=1}^{Q} \right)\\
\!\!&\le-\sum_{q=1}^{Q}\Big(\frac{\rho_q + 4^{q-1}\lambda_{\rm min}(\mathbf{H}^{H} \mathbf{H})}{2} - \frac{16^{q-1}\lambda_{\rm max}^2(\mathbf{H}^{H} \mathbf{H})}{\rho_q}\Big)\|\bar{\mathbf{x}}_q^{k+1}\!\!-\!\!\bar{\mathbf{x}}_q^k\|_2^2,
\end{split}
\end{equation}
where $\lambda_{\rm min}(\cdot)$ denotes the minimum igenvalues of a matrix.
\end{lemma}
\begin{proof}
We split LHS of the inequality \eqref{eq:L_difference} into three terms
\begin{equation}\label{eq:successive_L1}
\begin{split}
&L_{\rho_q}(\{\mathbf{z}_q^{k+1}, \bar{\mathbf{x}}_q^{k+1}, \mathbf{y}_q^{k+1}\}_{q=1}^Q)-L_{\rho_q}(\{\mathbf{z}^{k}_q, \bar{\mathbf{x}}_q^{k}, \mathbf{y}_q^{k}\}_{q=1}^Q)\\
&=\!\underbrace{\left(L_{\rho_q}(\{\mathbf{z}_q^{k+1}, \bar{\mathbf{x}}_q^{k+1},\! \mathbf{y}_q^{k+1}\}_{q=1}^Q)\!-\!L_{\rho_q}(\{\mathbf{z}_q^{k+1},\! \bar{\mathbf{x}}_q^{k+1},\! \mathbf{y}_q^{k}\}_{q=1}^Q)\right)}_{\rm{term\ 1}}\\
&\quad+\!\underbrace{\left(L_{\rho_q}(\{\mathbf{z}^{k+1}_q, \bar{\mathbf{x}}_q^{k}, \mathbf{y}_q^{k}\}_{q=1}^Q)-L_{\rho_q}(\{\mathbf{z}^{k}_q, \bar{\mathbf{x}}_q^{k}, \mathbf{y}_q^{k}\}_{q=1}^Q)\right)}_{\rm{term\ 2}}\\
&\quad+\!\underbrace{\left(L_{\rho_q}(\{\mathbf{z}^{k+1}_q, \bar{\mathbf{x}}_q^{k+1}, \mathbf{y}_q^{k}\}_{q=1}^Q)-L_{\rho_q}(\{\mathbf{z}^{k+1}_q, \bar{\mathbf{x}}_q^{k}, \mathbf{y}_q^{k}\}_{q=1}^Q)\right)}_{\rm{term\ 3}}. \nonumber
\end{split}
\end{equation}
For the first term, we have the following derivations
\begin{align}\label{eq:term1}
\begin{split}
&L_{\rho_q}(\{\mathbf{z}_q^{k+1}, \bar{\mathbf{x}}_q^{k+1}, \mathbf{y}_q^{k+1}\}_{q=1}^{Q})-L_{\rho_q}(\{\mathbf{z}_q^{k+1}, \bar{\mathbf{x}}_q^{k+1}, \mathbf{y}_q^{k}\}_{q=1}^{Q})\\
&\!=\!\sum_{q=1}^{Q}\Big(\big\langle \bar{\mathbf{x}}_q^{k+1}-\mathbf{z}_q^{k\!+\!1}, \mathbf{y}_q^{k\!+\!1}\big\rangle\!\!
-\!\!\big\langle \bar{\mathbf{x}}_q^{k\!+\!1}-\mathbf{z}_q^{k\!+\!1} , \mathbf{y}_q^{k}\big\rangle\Big) \\
&=\sum_{q=1}^{Q}\big\langle\bar{\mathbf{x}}_q^{k+1}-\mathbf{z}_q^{k+1},\; \mathbf{y}_q^{k+1}-\mathbf{y}_q^{k}\big\rangle \\
&\stackrel{\rm (a)}
=\sum_{q=1}^{Q}\frac{1}{\rho_q}\|\mathbf{y}_q^{k+1}-\mathbf{y}_q^{k}\|_2^2\\
&\stackrel{\rm (b)}
\le\sum_{q=1}^{Q}\frac{16^{q-1}\lambda_{\rm max}^2(\mathbf{H}^{H} \mathbf{H})}{\rho_q}\|\bar{\mathbf{x}}_q^{k+1}-\bar{\mathbf{x}}_q^{k}\|_2^2,
\end{split}
\end{align}
where ``$\stackrel{\rm (a)}{=}$'' and ``$\stackrel{\rm (b)}{\le}$'' comes from \eqref{y_q_update_L2Box} and \eqref{eq:z_difference} respectively.\\
For the second term, since $\mathbf{z}_q^{k+1}$ is the minimizer of $L_{\rho_q}(\{\mathbf{z}_q, \bar{\mathbf{x}}_q^{k}, \mathbf{y}_q^{k}\}_{q=1}^{Q})$, we obtain
\begin{equation}\label{eq:term2}
L_{\rho_q}(\{\mathbf{z}_q^{k+1}, \bar{\mathbf{x}}_q^{k}, \mathbf{y}_q^{k}\}_{q=1}^{Q})-L_{\rho_q}(\{\mathbf{z}_q^{k}, \bar{\mathbf{x}}_q^{k}, \mathbf{y}_q^{k}\}_{q=1}^{Q}) \le 0.
\end{equation}
For the third term, we have the following derivations
\begin{align}\label{eq:term3}
\begin{split}
&L_{\rho_q}(\{\mathbf{z}_q^{k+1}, \bar{\mathbf{x}}_q^{k+1}, \mathbf{y}_q^{k}\}_{q=1}^{Q})-L_{\rho_q}(\{\mathbf{z}_q^{k+1}, \bar{\mathbf{x}}_q^{k}, \mathbf{y}_q^{k}\}_{q=1}^{Q})\\
\le & \sum_{q=1}^{Q}\Bigl(\left\langle\nabla_{\bar{\mathbf{x}}_q}  L_{\rho_q}(\!\bar{\mathbf{x}}_1^{k+1}\!\!\!\!,\!\cdots\!,\bar{\mathbf{x}}_{q-1}^{k+1}\!,\!\bar{\mathbf{x}}_q^{k+1},\! \bar{\mathbf{x}}_{q+1}^k\!,\!\cdots\!,\!\bar{\mathbf{x}}_Q^k, \{ \mathbf{z}_q^{k+1},\mathbf{y}_q^{k}\}_{q=1}^{Q}), \bar{\mathbf{x}}_q^{k+1}-\bar{\mathbf{x}}_q^{k}\right\rangle\\
&\hspace{1.5cm}-\frac{\rho_q +4^{q-1}\lambda_{\rm min}(\mathbf{H}^{H} \mathbf{H})}{2}\|\bar{\mathbf{x}}_q^{k+1}-\bar{\mathbf{x}}^{k}_q\|_2^2 \Bigl)\\
 \le& -\!\!\sum_{q=1}^{Q}\frac{\rho_q +4^{q-1}\lambda_{\rm min}(\mathbf{H}^{H}\mathbf{H})}{2}\|\bar{\mathbf{x}}_q^{k+1}\!\!-\!\!\bar{\mathbf{x}}^{k}_q\|_2^2,
\end{split}
\end{align}
where the first inequality holds since $L_{\rho_q}(\{\mathbf{z}_q^{k+1}, \bar{\mathbf{x}}_q, \mathbf{y}_q^{k}\}_{q=1}^{Q})$ is strongly convex with respect to $\bar{\mathbf{x}}_q$ \cite{boyd2004convex} and the second inequality holds since $\bar{\mathbf{x}}_q^{k+1}$ is the minimizer of $L_{\rho_q}(\{\mathbf{z}_q^{k+1}, \bar{\mathbf{x}}_q, \mathbf{y}_q^{k}\}_{q=1}^{Q})$, i.e.,
\[
\begin{split}
\nabla_{\bar{\mathbf{x}}_q}  L_{\rho_q}(\bar{\mathbf{x}}_q^{k+1},\!\mathbf{x}_1^{k+1},\cdots,\mathbf{x}_{q-1}^{k+1}, \mathbf{x}_{q+1}^k,\cdots,\bar{\mathbf{x}}_Q^k,\{\mathbf{z}_q^{k+1}, \mathbf{y}_q^{k}\}_{q=1}^{Q})= 0.
\end{split}
\]
Adding both sides of inequalities \eqref{eq:term1}, \eqref{eq:term2}, and \eqref{eq:term3}, we can obtain
\begin{align*}
&L_{\rho_q}(\{\mathbf{z}_q^{k+1}, \bar{\mathbf{x}}_q^{k+1}, \mathbf{y}_q^{k+1}\}_{q=1}^{Q})-L_{\rho_q}(\{\mathbf{z}_q^{k}, \bar{\mathbf{x}}_q^{k}, \mathbf{y}_q^{k}\}_{q=1}^{Q})\\
&\le -\sum_{q=1}^{Q}\Big(\frac{\rho_q + 4^{q-1}\lambda_{\rm min}(\mathbf{H}^{H} \mathbf{H})}{2}-\frac{16^{q-1}\lambda_{\rm max}^2(\mathbf{H}^{H} \mathbf{H})}{\rho_q}\Big) \|\bar{\mathbf{x}}_q^{k+1}-\bar{\mathbf{x}}_q^{k}\|_2^2,
\end{align*}
which completes the proof.
\end{proof}

\begin{lemma}\label{lemma:L_bounded1}
Let $\rho_q > 4^{q-1}\sqrt{2} \lambda_{\rm max}(\mathbf{H}^H\mathbf{H})$. Assume tuples $\{\{\mathbf{z}_q^{k},\bar{\mathbf{x}}_q^k,\mathbf{y}_q^k\}_{q=1}^{Q}\}$ is generated by Algorithm 1, then $L_{\rho_q}(\{\mathbf{z}_q^{k},\bar{\mathbf{x}}_q^k,\mathbf{y}_q^k\}_{q=1}^{Q})$ is lower bounded as follows
\begin{align}\label{eq:Lbound}
\begin{split}
&L_{\rho_q}(\{\mathbf{z}_q^{k},\bar{\mathbf{x}}_q^k,\mathbf{y}_q^k\}_{q=1}^{Q}) \ge \mathit{f}\big(\mathbf{z}_q^{k}\big).
\end{split}
\end{align}
\end{lemma}
\begin{proof}
Plugging \eqref{eq:z_nabla_expression} into \eqref{eq:lagrangian_L2BoxADMM}, we obtain
\begin{align}\label{eq:Lbound1}
\begin{split}
L_{\rho_q}(\{\mathbf{z}_q^{k}, \bar{\mathbf{x}}_q^{k}, \mathbf{y}_q^{k}\}_{q=1}^{Q})=\mathit{f}\left(\bar{\mathbf{x}}_q^{k}\right)+\!\Big\langle \mathbf{z}_q^{k}\!-\!\bar{\mathbf{x}}_q^{k},\!\nabla_{\bar{\mathbf{x}}_q} \mathit{f}\left(\bar{\mathbf{x}}_q^{k}\right)\!\Big\rangle \!+\!\frac{\rho_q}{2}\Big\|\bar{\mathbf{x}}_q^{k}-\!\mathbf{z}_q^{k}\Big\|_2^2.\\
\end{split}
\end{align}
Since we show that gradient $\|\nabla_{\bar{\mathbf{x}}_q}\mathit{f}\left(\bar{\mathbf{x}}_q\right)\|_2$ is Lipschitz continuous in Lemma \ref{lemma:z1} and $ \|\nabla_{\bar{\mathbf{x}}_q}^2 \mathit{f}\left(\bar{\mathbf{x}}_q\right)\|_2 \le 4^{q-1}\lambda_{\rm max}(\mathbf{H}^H\mathbf{H})$, according to the Decent Lemma \cite{bertsekas1999nonlinear}, we can obtain
\begin{align*}
\mathit{f}\big(\mathbf{z}_q^{k}\big) &\!\le\! \mathit{f}\left(\bar{\mathbf{x}}_q^{k}\right)\!+\!\Big\langle \!\nabla_{\bar{\mathbf{x}}_q} \mathit{f}\left(\bar{\mathbf{x}}_q^{k}\right)\!, \mathbf{z}_q^{k}-\bar{\mathbf{x}}_q^{k} \!\Big\rangle+\frac{4^{q-1}\lambda_{\rm max}(\mathbf{H}^H\mathbf{H})}{2}\Big\|\mathbf{z}_q^{k}-\bar{\mathbf{x}}_q^{k}\Big\|_2^2,
\end{align*}
which can be further derived to the following inequality
\begin{align}\label{eq:Lbound2}
\begin{split}
&\mathit{f}\left(\bar{\mathbf{x}}_q^{k}\right)+\Big\langle \mathbf{z}_q^{k}-\bar{\mathbf{x}}_q^{k},\nabla_{\bar{\mathbf{x}}_q} \mathit{f}\left(\bar{\mathbf{x}}_q^{k}\right) \Big\rangle \ge \mathit{f}\big(\mathbf{z}_q^{k}\big)-\frac{4^{q-1}\lambda_{\rm max}(\mathbf{H}^H\mathbf{H})}{2}\Big\|\mathbf{z}_q^{k}-\bar{\mathbf{x}}_q^{k}\Big\|_2^2.
\end{split}
\end{align}
Plugging \eqref{eq:Lbound2} into \eqref{eq:Lbound1}, we can get
\begin{align}\label{eq:Lbound3}
\begin{split}
&L_{\rho_q}(\{\mathbf{z}_q^{k},\bar{\mathbf{x}}_q^k,\mathbf{y}_q^k\}_{q=1}^{Q}) \ge \mathit{f}\big(\mathbf{z}_q^{k}\big)+\frac{\rho_q-4^{q-1}\lambda_{\rm max}(\mathbf{H}^H\mathbf{H})}{2}\Big\|\bar{\mathbf{x}}_q^{k}-\mathbf{z}_q^{k}\Big\|_2^2.
\end{split}
\end{align}
Since $\mathit{f}\big(\mathbf{z}_q^{k}\big)$ is bounded over $\mathbf{z}_{1q}\in\mathbb{S}_b,\mathbf{z}_{2q}\in\mathbb{S}_2$, as well as the fact that $\rho_q-4^{q-1}\lambda_{\rm max}(\mathbf{H}^H\mathbf{H})>0$ comes from $\rho_q>4^{q-1} \sqrt{2} \lambda_{\rm max}(\mathbf{H}^H\mathbf{H})$. Using these two cases leads to the desired result that $L_{\rho_q}(\{\mathbf{z}_q^{k},\bar{\mathbf{x}}_q^k,\mathbf{y}_q^k\}_{q=1}^{Q})$ is lower bounded and Lemma \ref{lemma:L_bounded1} has been proved.
 \end{proof}

According to Lemma 2, summing both sides of the inequality \eqref{eq:L_difference} when $k=1,2,\cdots,+\infty$, we can obtain
\begin{equation}
    \begin{split}
    &L_{\rho_q}\left (\{\mathbf{z}_q^{1}, \bar{\mathbf{x}}_q^{1}, \mathbf{y}_q^{1} \}_{q=1}^{Q}\right)-\lim_{k\rightarrow +\infty}L_{\rho_q}\left (\{\mathbf{z}_q^{k}, \bar{\mathbf{x}}_q^{k}, \mathbf{y}_q^{k}\}_{q=1}^{Q}\right)\\
      &\geq \sum_{k=1}^{+\infty}\sum_{q=1}^{Q}\Big(\frac{\rho_q + 4^{q-1}\lambda_{\rm min}(\mathbf{H}^{H} \mathbf{H})}{2}-\frac{16^{q-1}\lambda_{\rm max}^2(\mathbf{H}^{H} \mathbf{H})}{\rho_q}\Big)\|\bar{\mathbf{x}}_q^{k+1}-\bar{\mathbf{x}}_q^k\|_2^2.\\
  \end{split}
  \end{equation}
From Lemma 3, one can see that $\lim_{k\rightarrow +\infty}L_{\rho_q}\left (\{\mathbf{z}_q^{k}, \bar{\mathbf{x}}_q^{k}, \mathbf{y}_q^{k}\}_{q=1}^{Q}\right)>-\infty$. Moreover, since $\frac{\rho_q + 4^{q-1}\lambda_{\rm min}(\mathbf{H}^{H} \mathbf{H})}{2}-\frac{16^{q-1}\lambda_{\rm max}^2(\mathbf{H}^{H} \mathbf{H})}{\rho_q}\geq0$, we can obtain
\begin{equation}
 \lim\limits_{k\rightarrow+\infty} \|\bar{\mathbf{x}}_q^{k+1}-\bar{\mathbf{x}}_q^k\|_2 = 0. \label{eq:differenceX0}
\end{equation}
Plugging \eqref{eq:differenceX0} into RHS of equation \eqref{eq:z_difference}, we get
\begin{equation}\label{limYzero}
\lim\limits_{k\rightarrow+\infty}\|\mathbf{y}_q^{k+1}-\mathbf{y}_q^k\|_2 = 0.
\end{equation}
Plugging \eqref{limYzero} into \eqref{y_q_update_L2Box}, we get
\begin{align}\label{eq:differenceX01}
\lim\limits_{k\rightarrow+\infty}\|\bar{\mathbf{x}}_q^{k+1}-\mathbf{z}_q^{k+1} \|_2 = 0.
\end{align}
Combining \eqref{eq:differenceX0} and \eqref{eq:differenceX01}, we obtain that
\begin{align}\label{eq:differenceYq}
\lim\limits_{k\rightarrow+\infty}\|\mathbf{z}_q^{k+1}-\mathbf{z}_q^{k} \|_2 = 0.
\end{align}
Since $\mathbf{z}_{1q}\in\mathbb{S}_b,\mathbf{z}_{2q}\in\mathbb{S}_2$ are bounded, we can obtain the following convergence results from \eqref{eq:differenceYq}.
\begin{equation}\label{convergence yq}
\lim\limits_{k\rightarrow+\infty}\mathbf{z}_q^{k}\!=\!\mathbf{z}_q^{*}.
\end{equation}
Plugging \eqref{convergence yq} into \eqref{eq:differenceX01}, we can see
\begin{equation}\label{convergence_x0}
\lim\limits_{k\rightarrow+\infty}\bar{\mathbf{x}}_q^{k}=\bar{\mathbf{x}}_q^{*}=\mathbf{z}_q^{*}.
 \end{equation}
From \eqref{eq:z_nabla_expression},  we can derive
  \begin{equation}\label{limy_x0}
    \lim\limits_{k\rightarrow+\infty} \mathbf{y}_q^k = \lim\limits_{k\rightarrow+\infty}-\nabla_{\bar{\mathbf{x}}_q} \mathit{f}\left(\bar{\mathbf{x}}_q^{k}\right).
   \end{equation}
Since $\|\nabla_{\bar{\mathbf{x}}_q} \mathit{f}\left(\bar{\mathbf{x}}_q^{k+1}\right)-\nabla_{\bar{\mathbf{x}}_q} \mathit{f}\left(\bar{\mathbf{x}}_q^{k}\right)\|_2^2 \le 16^{q-1}\lambda_{\rm max}^2(\mathbf{H}^{H} \mathbf{H}) \|\bar{\mathbf{x}}_q^{k+1}-\bar{\mathbf{x}}_q^{k}\|_2^2$ and $\bar{\mathbf{x}}_q$ is bounded, we can conclude that all the elements in $\nabla_{\bar{\mathbf{x}}_q} \mathit{f}\left(\bar{\mathbf{x}}_q\right)$ are also bounded. Therefore, equation \eqref{limYzero} indicates
\begin{equation}\label{convergence z}
\lim\limits_{k\rightarrow+\infty} \mathbf{y}_q^k =  \mathbf{y}_q^{*}.
\end{equation}

Then, we prove $(\{\mathbf{z}_q^*, \bar{\mathbf{x}}_q^*,\mathbf{y}_q^*\}_{q=1}^{Q}\})$ is a stationary point of problem \eqref{eq: L2-Box ML}.

We discuss the two variables $\mathbf{z}_{1q}$ and $\mathbf{z}_{2q}$ in $\{\mathbf{z}_q^{k+1}\}_{q=1}^{Q} = \underset{\mathbf{z}_{1q}\in\mathbb{S}_b,\mathbf{z}_{2q}\in\mathbb{S}_2}{\arg\min} \;  L_{\rho_q}(\{\mathbf{z}_q,\bar{\mathbf{x}}_q^k, \mathbf{y}_q^k)$ (see \eqref{z_q_update_L2Box}) respectively as follows
\begin{subequations}\label{L2BoxADMM_update_respec}
\begin{align}
&\mathbf{z}_{1q}^{k+1} = \mathop{\arg\min}_{\mathbf{z}_{1q}\in\mathbb{S}_b}  L_{\rho_{1q},\rho_{2q}}(\{\mathbf{z}_{1q},\mathbf{x}_q^k, \mathbf{y}_{1q}^k\}_{q=1}^{Q}), \label{z_1q_update_L2Box}\\
&\mathbf{z}_{2q}^{k+1} = \mathop{\arg\min}_{\mathbf{z}_{2q}\in\mathbb{S}_2}  L_{\rho_{1q},\rho_{2q}}(\{\mathbf{z}_{2q},\mathbf{x}_q^k, \mathbf{y}_{2q}^k\}_{q=1}^{Q}), \label{z_2q_update_L2Box}
\end{align}
\end{subequations}

First, from \eqref{z_1q_update_L2Box} and $\mathbb{S}_b$ is convex, we have the following optimality conditions
\vspace{-2pt}

\begin{equation}\label{optimality_zq}
\begin{split}
&\Big\langle\nabla_{\mathbf{z}_{1q}}\Big(\mathit{f}\left(\mathbf{x}_q^k\right) + \textstyle\sum\nolimits_{q=1}^{Q}\big\langle \mathbf{x}_q^k - \mathbf{z}_{1q}^{k+1}, \mathbf{y}_{1q}^k\big\rangle  + \sum\nolimits_{q=1}^{Q}\frac{\rho_{1q}}{2}\big\|\mathbf{x}_q^k - \mathbf{z}_{1q}^{k+1}\big\|_2^2\Big),\mathbf{z}_{1q}-\mathbf{z}_{1q}^{k+1} \Big\rangle \ge 0,\\
&\hspace{9.1cm}   \quad \forall~\mathbf{z}_{1q}\in\mathbb{S}_b, ~~ q=1,2,\cdots,Q.
\end{split}
\end{equation}
It can be further obtained
\begin{equation}\label{optimality_zq_simple}
\begin{split}
&\Big\langle-\mathbf{y}_{1q}^k + \rho_{1q}(\mathbf{z}_{1q}^{k+1}-\mathbf{x}_q^k),~\mathbf{z}_{1q}-\mathbf{z}_{1q}^{k+1} \Big\rangle \ge 0,\\
&\hspace{2cm}   \quad \forall~\mathbf{z}_{1q}\in\mathbb{S}_b,~ q=1,2,\cdots,Q.
\end{split}
\end{equation}
When $k\rightarrow+\infty$, plugging convergence results \eqref{convergence yq}, \eqref{convergence_x0}, and \eqref{convergence z} into \eqref{optimality_zq_simple}, it can be simplified as

\begin{equation}
\Big \langle\mathbf{z}_{1q}-\mathbf{z}^*_{1q}, -\mathbf{y}_{1q}^*\Big\rangle \ge 0.
\end{equation}

Second, to be clear, we rewrite \eqref{z_2q_update_L2Box} as
\begin{subequations}\label{neq:z_2q}
\begin{align}
&\hspace{0.4cm}\min_{\mathbf{z}_{2q}} \hspace{0.2cm} \frac{\rho_{2q}}{2}\big\|\mathbf{x}^k_q\!-\!\mathbf{z}_{2q}\big\|_2^2+\mathbf{y}_{2q}^{kT}\big(\mathbf{x}^k_q\!-\!\mathbf{z}_{2q}\big),  \label{neq:z_2qa} \\
&\hspace{0.5cm}{\rm {s.t.}} \hspace{0.3cm} \|\mathbf{z}_{2q}\|_2^2=2U, \label{neq:z_2qb}
\end{align}
\end{subequations}
which can be further equivalent to
\begin{subequations}\label{neq:z_2q2}
\begin{align}
&\hspace{0.4cm}\min_{\mathbf{z}_{2q}} \hspace{0.2cm} -(\rho_{2q}\mathbf{x}^{k}_q+\mathbf{y}^k_{2q})^T\mathbf{z}_{2q},  \label{neq:z_2qa2} \\
&\hspace{0.5cm}{\rm {s.t.}} \hspace{0.3cm} \|\mathbf{z}_{2q}\|_2^2=2U. \label{neq:z_2qb2}
\end{align}
\end{subequations}
It is easy to see that its optimal solution $\mathbf{z}_{2q}^{k+1}$ should satisfy
\begin{equation}\label{neq:z_2q3}
\mathbf{z}_{2q}^{k+1}=c(\rho_{2q}\mathbf{x}^{k}_q+\mathbf{y}^k_{2q}),
\end{equation}
where $c$ is a positive constant. Plug \eqref{neq:z_2q3} into $ \|\mathbf{z}_{2q}\|_2^2=2U$. Noticing $c>0$, we can determine
\begin{equation}\label{c}
  c = \frac{\sqrt{2U}}{\|\rho_{2q}\mathbf{x}^{k}_q+\mathbf{y}^k_{2q}\|_2},
\end{equation}
which results in
\begin{equation}\label{z_2q_k+1_2}
  \mathbf{z}_{2q}^{k+1}=\frac{\sqrt{2U}(\rho_{2q}\mathbf{x}^k_q+\mathbf{y}^k_{2q})}{\lVert\rho_{2q}\mathbf{x}^k_q+\mathbf{y}^k_{2q}\rVert_2}.
\end{equation}

When $k\rightarrow+\infty$, plugging convergence results \eqref{convergence yq}, \eqref{convergence_x0}, and \eqref{convergence z} into \eqref{z_2q_k+1_2}, the optimal solution $\mathbf{z}_{2q}$ is get
\begin{equation}\label{z_2q_optimal}
  \mathbf{z}_{2q}^*=\frac{\sqrt{2U}(\rho_{2q}\mathbf{x}^*_q+\mathbf{y}^*_{2q})}{\lVert\rho_{2q}\mathbf{x}^*_q+\mathbf{y}^*_{2q}\rVert_2}.
 \end{equation}

Combining \eqref{convergence_x0}, \eqref{limy_x0}, and \eqref{convergence z}, the following conditions hold
\begin{equation}\label{stationary point2}
\mathbf{y}_q^*=-\nabla_{\mathbf{x}_q} \mathit{f}\left(\bar{\mathbf{x}}_q^*\right),\ \ \bar{\mathbf{x}}_q^* = \mathbf{z}_{q}^*.
\end{equation}
It concludes the proof of Theorem \ref{thm:convergence}.
\end{proof}
\section{Iteration Complexity Analysis}
\label{sec:analysis-complexity}

\begin{theorem}\label{iteration complexity}
  Let $t$ be the minimum iteration index such that the residual $\sum_{q=1}^{Q}\|\bar{\mathbf{x}}_q^{k+1}-\bar{\mathbf{x}}_q^{k}\|_2^2<\epsilon$, where $\epsilon$ is a desired precise parameter for the solution. Then, we have the following iteration complexity result
 \[
   \begin{split}
     t \!\leq \!\frac{1}{C\epsilon}\Big(\!L_{\rho_q}(\{\mathbf{z}^{1}_q, \bar{\mathbf{x}}_q^{1}, \mathbf{y}_q^{1}\}_{q=1}^{Q}) -\mathit{f}\left(\bar{\mathbf{x}}_q^*\right) \Big),
   \end{split}
 \]
where the constant $ C=\min\left\{ \frac{\rho_q + 4^{q-1}\lambda_{\rm min}(\mathbf{H}^{H} \mathbf{H})}{2}-\frac{16^{q-1}\lambda_{\rm max}^2(\mathbf{H}^{H} \mathbf{H})}{\rho_q}\right\}_{q=1}^{Q}$.
\end{theorem}
{\it Remarks:}
Theorem \ref{iteration complexity} indicates that the ``{\it minimum iterations}'' is upper-bounded for the given precise $\epsilon$, which gives a low bound for the computational complexity of the proposed $\ell_2$-box ADMM algorithm.

\begin{proof}
To be clear, here we rewrite \eqref{eq:L_difference} as
\[
  \begin{split}
    &L_{\rho_q}\left (\{\mathbf{z}_q^{k}, \bar{\mathbf{x}}_q^{k}, \mathbf{y}_q^{k} \}_{q=1}^{Q}\right)-L_{\rho_q}\left (\{\mathbf{z}_q^{k+1}, \bar{\mathbf{x}}_q^{k+1}, \mathbf{y}_q^{k+1}\}_{q=1}^{Q}\right)\\
      &\geq \sum_{q=1}^{Q}\Big(\frac{\rho_q + 4^{q-1}\lambda_{\rm min}(\mathbf{H}^{H} \mathbf{H})}{2}-\frac{16^{q-1}\lambda_{\rm max}^2(\mathbf{H}^{H} \mathbf{H})}{\rho_q}\Big)\|\bar{\mathbf{x}}_q^{k+1}-\bar{\mathbf{x}}_q^k\|_2^2.\\
  \end{split}
\]

According to Lemma \ref{lemma:L_difference}, there exists a constant $C=\min\left\{ \frac{\rho_q + 4^{q-1}\lambda_{\rm min}(\mathbf{H}^{H} \mathbf{H})}{2}-\frac{16^{q-1}\lambda_{\rm max}^2(\mathbf{H}^{H} \mathbf{H})}{\rho_q}\right\}_{q=1}^{Q}$
such that
\[
  \begin{split}
    &L_{\rho_q}\left (\{\mathbf{z}_q^{k}, \bar{\mathbf{x}}_q^{k}, \mathbf{y}_q^{k}\}_{q=1}^{Q} \right)-L_{\rho_q}\left (\{\mathbf{z}_q^{k+1}, \bar{\mathbf{x}}_q^{k+1}, \mathbf{y}_q^{k+1}\}_{q=1}^{Q}\right)\\
      &\geq C\sum_{q=1}^{Q}\|\bar{\mathbf{x}}_q^{k+1}-\bar{\mathbf{x}}_q^k\|_2^2.\\
  \end{split}
\]
 Summing both sides of the above inequality from $k=1,\cdots, T$, we have
 \begin{equation}\label{dL}
   \begin{split}
      &L_{\rho_q}\left (\{\mathbf{z}_q^{1}, \bar{\mathbf{x}}_q^{1}, \mathbf{y}_q^{1} \}_{q=1}^{Q}\right)-L_{\rho_q}\left (\{\mathbf{z}_q^{T+1}, \bar{\mathbf{x}}_q^{T+1}, \mathbf{y}_q^{T+1}\}_{q=1}^{Q}\right)\\
      &\geq \sum_{k=1}^{T}\Big(C\sum_{q=1}^{Q}\|\bar{\mathbf{x}}_q^{k+1}-\bar{\mathbf{x}}_q^k\|_2^2\Big).\\
  \end{split}
 \end{equation}
 Since $t = \underset{k}{\rm min}\{k|\sum_{q=1}^{Q}\|\bar{\mathbf{x}}_q^{k+1}-\bar{\mathbf{x}}_q^{k}\|_2^2\leq\epsilon\}$, we can change \eqref{dL} to
 \begin{equation}\label{Relax L}
   \begin{split}
    & L_{\rho_q}\left (\{\mathbf{z}_q^{1}, \bar{\mathbf{x}}_q^{1}, \mathbf{y}_q^{1}\}_{q=1}^{Q} \right)-L_{\rho_q}\left (\{\mathbf{z}_q^{T+1}, \bar{\mathbf{x}}_q^{T+1}, \mathbf{y}_q^{T+1}\}_{q=1}^{Q}\right) \geq tC\epsilon.
   \end{split}
 \end{equation}
 Since we have $L_{\rho_q} \!\left (\{\mathbf{z}_q^{T+1}, \bar{\mathbf{x}}_q^{T+1}, \mathbf{y}_q^{T+1}\}_{q=1}^{Q}\right)\! \geq\! L_{\rho_q}\! \left (\{\mathbf{z}_q^*, \bar{\mathbf{x}}_q^*, \mathbf{y}_q^*\}_{q=1}^{Q}\right)$, \eqref{Relax L} can be reduced to
 \[
   \begin{split}
     t \!&\leq\! \frac{1}{C\epsilon}\bigg(\!L_{\rho_q}\left (\{\mathbf{z}_q^{1}, \bar{\mathbf{x}}_q^{1}, \mathbf{y}_q^{1} \}_{q=1}^{Q}\right) \!-\! L_{\rho_q}\left (\{\mathbf{z}_q^*, \bar{\mathbf{x}}_q^*, \mathbf{y}_q^*\}_{q=1}^{Q}\right)\!\bigg),
   \end{split}
 \]
 where $L_{\rho_q}\left (\{\mathbf{z}_q^*, \bar{\mathbf{x}}_q^*, \mathbf{z}^*\}_{q=1}^{Q}\right)=\mathit{f}\left(\bar{\mathbf{x}}_q^*\right)$. It concludes the proof of Theorem \ref{iteration complexity}.
\end{proof}

\ifCLASSOPTIONcaptionsoff
  \newpage
\fi